\documentclass[conference,a4paper,]{IEEEtran}

\usepackage[T1]{fontenc}
\usepackage[latin1]{inputenc}
\usepackage{graphicx}
\usepackage{theorem}
\usepackage{amsmath}
\usepackage{amsfonts}
\usepackage[caption=false,font=footnotesize]{subfig}
\usepackage[acronym,shortcuts]{glossaries}

\makeglossaries

\newacronym{QC}{QC}{quasi-cyclic}
\newacronym{QC-LDPC}{QC-LDPC}{quasi-cyclic low-density parity-check}
\newacronym{LDPC}{LDPC}{low-density parity-check}
\newacronym{LDPCC}{LDPCC}{low-density parity-check convolutional}
\newacronym{AC-LDPC}{AC-LDPC}{array convolutional low-density parity-check}
\newacronym{PDC-LDPC}{PDC-LDPC}{progressive differences convolutional low-density parity-check}
\newacronym{SC-LDPC}{SC-LDPC}{spatially coupled low-density parity-check}
\newacronym{AWGN}{AWGN}{additive white Gaussian noise}
\newacronym{BER}{BER}{bit error rate}
\newacronym{FER}{FER}{frame error rate}
\newacronym{TUB}{TUB}{truncated union bound}
\newacronym{BPSK}{BPSK}{binary phase shift keying}
\newacronym{SPA-LLR}{SPA-LLR}{sum-product algorithm with log-likelihood ratios}
\newacronym{RTI}{RTI}{regular time-invariant}
\newacronym{RTI-LDPCC}{RTI-LDPCC}{regular time-invariant low-density parity-check convolutional}

{\theorembodyfont{\rmfamily}
   }
{\theorembodyfont{\rmfamily}
   \newtheorem{Pro}{{\textbf Proposition}}[section]}
{\theorembodyfont{\rmfamily}
   \newtheorem{Lem}{{\textbf Lemma}}[section]}
{\theorembodyfont{\rmfamily}
   }
{\theorembodyfont{\rmfamily}
   }
{\theorembodyfont{\rmfamily}
   }
 
\def\HH{\mathbf{H}}
\def\0{\mathbf{0}}
\def\Hconv{\mathbf{H}}

\begin{document}

\title{Time-Invariant Spatially Coupled Low-Density Parity-Check Codes with Small Constraint Length}

\author{\IEEEauthorblockN{Marco Baldi, Massimo Battaglioni, Franco Chiaraluce and Giovanni Cancellieri}
\IEEEauthorblockA{Dipartimento di Ingegneria dell'Informazione\\
Universit\`a Politecnica delle Marche\\
Ancona, Italy\\
Email:  \{m.baldi, f.chiaraluce, g.cancellieri\}@univpm.it, m.battaglioni@pm.univpm.it}}

\maketitle
\begin{abstract}
We consider a special family of \ac{SC-LDPC} codes, that is,
time-invariant \ac{LDPCC} codes, which are known in the literature for a long time.
Codes of this kind are usually designed by starting from \ac{QC} block codes, 
and applying suitable unwrapping procedures.
We show that, by directly designing the \ac{LDPCC} code syndrome former matrix 
without the constraints of the underlying \ac{QC} block code, it is possible to achieve 
smaller constraint lengths with respect to the best solutions available in the literature.
We also find theoretical lower bounds on the syndrome former constraint length
for codes with a specified minimum length of the local cycles in their Tanner graphs.
For this purpose, we exploit a new approach based on a numerical representation
of the syndrome former matrix, which generalizes over a technique we already
used to study a special subclass of the codes here considered.
\end{abstract}


\section{Introduction}

\glsresetall

\Ac{SC-LDPC} codes represent a cutting-edge innovation in the context of modern channel coding in general and of \ac{LDPC} coding in particular.
In fact, classical \ac{LDPC} block codes \cite{Gallager} are known to approach the channel capacity under belief propagation decoding \cite{Richardson2001}.
\ac{SC-LDPC} codes represent a further step in this direction, since they are able to further reduce the gap to capacity \cite{Kudekar2013} thanks to the 
threshold saturation phenomenon.

A special class of \ac{SC-LDPC} codes is that of \ac{LDPCC} codes, which have been shown to outperform their block counterparts \cite{Tanner2004}.
These codes are usually designed by starting from \ac{QC-LDPC} codes \cite{Lin2004Book} and using a technique known as \textit{unwrapping} to produce 
a semi-infinite description of the convolutional code \cite{Tanner2004, Felstrom1999, Pusane2011}.
This approach has allowed to design \ac{LDPCC} codes with very good performance \cite{Bocharova2012, Smarandache2012}.
However, despite some attempts to achieve small constraint lengths have been done \cite{Baldi2014}, starting from \ac{QC-LDPC} codes and then unwrapping
them usually results in \ac{LDPCC} codes with large constraint lengths.

In fact, shift register-based circuits like that in \cite[Fig. 4]{Felstrom1999} can be used to perform encoding of an \ac{LDPCC} code, while decoding can be
performed through iterative message passing algorithms working on a window sliding over the received sequence \cite{Felstrom1999}.
Complexity of these encoding and decoding techniques increases linearly with the syndrome former constraint length of the code.
Therefore, designing codes with small constraint length is a valuable target from the complexity standpoint.

In this paper we study the design of time-invariant \ac{LDPCC} codes without starting from \ac{QC-LDPC} block codes.
This is done by directly designing the syndrome former matrix which then forms the semi-infinite parity-check matrix of the \ac{LDPCC} code.
We follow an approach similar to that proposed in \cite{Baldi2012b}, where we introduced a special class of \ac{LDPCC} codes named 
\ac{PDC-LDPC} codes.
The codes considered in \cite{Baldi2012b} have rate $\frac{a-1}{a}$, with $a$ being an integer $> 1$, and local cycles with length $\ge 6$
in their associated Tanner graph. Another solution to design codes with the same parameters has been proposed in \cite{Cho2015}.
Here we generalize the approach proposed in \cite{Baldi2012b} to the design of codes with rate $\frac{a-c}{a}$, with $a$ and $c$ being two positive
integers such that $a > c$, and minimum length $g$ of the local cycles in their Tanner graphs.
The numerical representation we adopt for the syndrome former matrix significantly facilitates searching for short cycles in the code 
Tanner graph. Similar efficient searches have recently been performed for \ac{QC-LDPC} block codes \cite{Tasdighi2016}.
This approach permits us to perform theoretical and exhaustive analyses for $g=6$ and $g=8$, as well as Montecarlo assessments for larger values of $g$. 

The organization of the paper is as follows.
In Section \ref{sec:TICodes} we remind the definition of time-invariant \ac{SC-LDPC} codes and their relevant parameters.
In Section \ref{sec:LocalCycles} we introduce a numerical description of the syndrome former matrix which facilitates the search of local cycles.
In Section \ref{sec:MinConstLength} we provide theoretical bounds on the minimum constraint length which is needed to avoid local cycles up to a given length.
In Section \ref{sec:Examples} we provide a comparative assessment of the bounds with exhaustive searches as well as some results based on Montecarlo simulations.
Section \ref{sec:Conclusion} concludes the paper.

\section{Time-invariant spatially coupled low-density parity-check codes \label{sec:TICodes} }

The codes we consider are defined by semi-infinite parity-check matrices in the form \eqref{eq:Hconv},
where each block $\HH_i$, $i = 0, 1, 2, \ldots, m_h$, is a binary matrix with size $c \times a$.
The syndrome former matrix $\HH_s = \left[ \HH_0^T | \HH_1^T | \HH_2^T | \ldots | \HH_{m_h}^T \right]$, where $^T$ denotes transposition, has $a$ rows and $L_h$ columns. As evident in \eqref{eq:Hconv}, $\HH$ is obtained by $\HH_s^T$ and its replicas, shifted one each other by $c$ positions. The time invariant \ac{LDPCC} code defined by \eqref{eq:Hconv} has asymptotic code rate $R = \frac{a-c}{a}$,
syndrome former memory order $m_h = \left\lceil \frac{L_h}{c} \right\rceil - 1$
and syndrome former constraint length $v_s = (m_h + 1) a = \left\lceil \frac{L_h}{c} \right\rceil a$.

\begin{equation}
\HH = \left[\begin{array}{cccccc}
\arraycolsep=1.4pt\def\arraystretch{5pt}
\HH_0 				& \0 							& \0 							& \ddots \\
\HH_1				& \HH_0					& \0 							& \ddots \\
\HH_2				& \HH_1					& \HH_0					& \ddots \\
\vdots 			& \HH_2 				& \HH_1 				& \ddots \\
\HH_{m_h}	& \vdots 				& \HH_2					& \ddots \\
\0 						& \HH_{m_h}	& \vdots					& \ddots \\
\0 						& \0 							& \HH_{m_h}	& \ddots \\
\0 						& \0 							& \0							& \ddots \\
\vdots				& \vdots					& \vdots					& \ddots \\
\end{array}\right].
\label{eq:Hconv}
\end{equation}

An alternative representation of $\HH_s$ which is often used in the literature exploits polynomials $\in F_2[x]$. In this case, the code is described by a $c \times a$ matrix with polynomial entries, 
that is
\begin{equation}
H(x)=\left[\begin{array}{llll}
h_{0,0}(x) & h_{0,1}(x) & \ldots & h_{0,a-1}(x)\\
h_{1,0}(x) & h_{1,1}(x) & \ldots & h_{1,a-1}(x)\\
\vdots & \vdots & \ddots & \vdots\\
h_{c-1,0}(x) & h_{c-1,1}(x) & \ldots & h_{c-1,a-1}(x)\end{array}\right],
\label{eq:Hx}
\end{equation}
where each $h_{i,j}(x)$, $i = 0, 1, 2, \ldots, c-1$, $j = 0, 1, 2, \ldots, a-1$, is a polynomial $\in F_2[x]$ or a null term.
The code representation based on $\HH_s$ can be converted into that based on $H(x)$ through the following simple procedure.
First of all, starting from $\HH_s$, the multiset $I$ containing the sets of indexes (beginning from zero) of the symbols $1$ 
in each row of $\HH_s$ must be computed.
Then, the $j$-th column of $H(x)$ is obtained from the set $I_j \in I$, $j =0, 1, 2, \ldots, a-1$, as follows:
\begin{enumerate}
\item Initialize $h_{i,j}(x) = 0$, $i = 0, 1, 2, \ldots, c-1$.
\item $\forall l \in I_j$, compute $l_d = \left\lfloor l/c \right\rfloor$, $l_m = l \mod c$ and add $x^{l_d}$ to $h_{l_m,j}(x)$.
\end{enumerate}

This procedure is an inverse of the unwrapping techniques proposed in \cite{Tanner2004, Felstrom1999}.
In fact, most previous works are devoted to the design of $H(x)$ and then $\HH_s$ is obtained through unwrapping.
However, designing $H(x)$ requires to first choose the form of the polynomials $h_{i,j}(x)$ (null, monomials, binomials, etc.) and then optimize their exponents.
Such an approach has also been followed in \cite{Baldi2015}, where some low rate \ac{LDPCC} codes with small constraint length have been found.
The matrix $H(x)$ is also used in \cite{Zhou2012} to find unavoidable cycles and design \ac{LDPCC} codes free of short local loops.
In this paper we aim at finding the codes with minimum constraint length over all possible configurations.
For this purpose, working with $\HH_s$ is advantageous in that it allows to perform a single step optimization over all possible choices.
Therefore, we focus on $\HH_s$ and we need the transformation from $\HH_s$ to $H(x)$ described above to perform comparisons with the design examples reported in previous works.
As we will see in Section \ref{sec:Examples}, our approach allows to find codes with shorter constraint length than those in \cite{Zhou2012}.

\section{Local cycles \label{sec:LocalCycles}}

Local cycles are closed loops starting from a node of the Tanner graph associated to an \ac{LDPC} code and returning to the same node by passing
only once through any edge.
Since the Tanner graph is derived from the code parity-check matrix, local cycles can be defined over such a matrix as well.
This way, we are able to directly relate the constraint length of an \ac{SC-LDPC} code to its local cycles length.

Following an approach similar to that introduced in \cite{Baldi2012b}, we describe the matrix $\HH_s$ through a set of integer values
representing the differences between each pair of ones in each row of $\HH_s$.
These differences are denoted as $\delta_{i,j}$, where $i$ is the row of $\HH_s$ $(i=0,1,2,\ldots,a-1)$ 
and $j$ is the column of $\HH_s$ corresponding to the first of the two symbols $1$ forming the difference $(j=0,1,2,\ldots, L_h-2)$.
The index of the second symbol $1$ forming the difference is easily found as $j + \delta_{i,j}$.
For each difference we also compute the values of two \textit{levels} which are relative to the value of the parameter $c$.
The \textit{starting level} is defined as $l_s = j \mod c$, while the \textit{ending level} is defined as $l_e = (j + \delta_{i,j}) \mod c$.
Both levels obviously take values in $\left\{0,1,2 \ldots, c-1\right\}$.

\begin{figure}[!t]
\begin{centering}
\includegraphics[width=70mm,keepaspectratio]{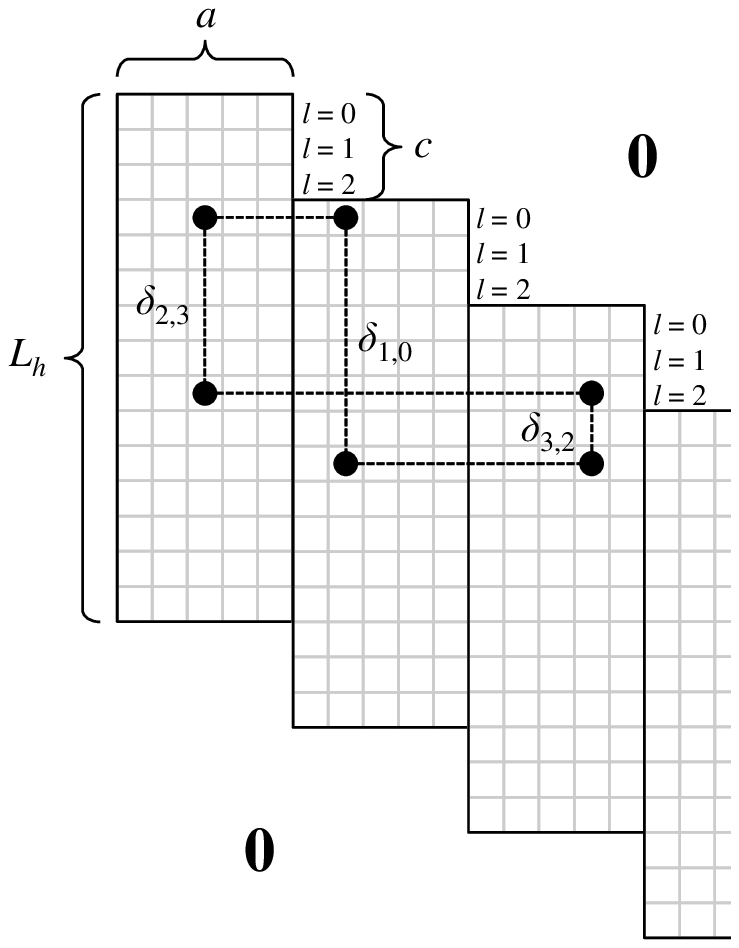}
\caption{Example of $\Hconv$ with a local cycle of length $6$.}
\label{fig:CycleExample}
\par\end{centering}
\end{figure}

Based on this representation of the syndrome former matrix, it is easy to identify closed loops in the Tanner graph associated to $\Hconv$.
In fact, a local cycle occurs every time a sum of the type $\delta_{i_1,j_1} \pm \delta_{i_2,j_2} \pm \ldots \pm \delta_{i_l,j_l}$ equals zero,
and the length of the cycle is $2l$, with $l$ being an integer $> 1$. 
An example is reported in Fig. \ref{fig:CycleExample}, where a cycle with length $6$ corresponds to the relation $\delta_{2,3} + \delta_{3,2} - \delta_{1,0} = 0$.
Not all the possible sums or differences of $\delta_{i,j}$ are valid to generate local cycles.
In fact, $\delta_{x,y}$ can be added to $\delta_{i,j}$ iff the starting level of the former coincides with the ending level of the latter.
Instead, $\delta_{x,y}$ can be subtracted to $\delta_{i,j}$ iff their ending levels coincide.
In addition, the first and the last levels of the sum $\delta_{i_1,j_1} \pm \delta_{i_2,j_2} \pm \ldots \pm \delta_{i_l,j_l}$ must coincide.
Let us denote as $\left. \delta_{i,j} \right._{(l_e)}^{(l_s)}$ the difference $\delta_{i,j}$ with its associated starting and ending levels.
For the example reported in Fig. \ref{fig:CycleExample}, we have $\left. \delta_{2,3} \right._{(2)}^{(0)} + \left. \delta_{3,2} \right._{(1)}^{(2)} - \left. \delta_{1,0} \right._{(1)}^{(0)} = 0$,
which therefore complies with the above rules.
Owing to the special structure of $\Hconv$, some further rules hold concerning the existence of closed loops.
In fact, it must be taken into account that the shift of the replicas of $\HH_s^T$ within $\Hconv$ is neither cyclic nor quasi-cyclic.
Therefore, a closed loop due to the differences in a single row of $\HH_s$ can or cannot exist depending on the positions of the symbols $1$ in that row.
For example, a cycle with length $6$ due to a single row of $\HH_s$ with weight $w \ge 3$ exists iff at least $3$ symbols $1$ are at the same level.
Instead, a circulant matrix with row weight $\ge 3$ always yields length $6$ cycles.
Moreover, in a sum of differences, the same $\delta_{i,j}$ cannot appear with both signs in two adjacent terms. Based on these considerations, for a given matrix $\HH_s$ a very efficient numerical procedure can be exploited to find all the local
cycles with a given maximum length.
Such a procedure has been implemented in software, and has allowed to perform exhaustive (when possible) or Montecarlo (otherwise) 
analyses of the syndrome former matrices with minimum constraint length and free of local cycles up to a given size.
Moreover, by studying the cases in which differences may or may not be summed or subtracted, it is possible to obtain lower bounds
on the minimum constraint length which is needed to avoid local cycles up to a given length, as described in the next
section.

\section{Minimum constraint length \label{sec:MinConstLength} }

Let us consider some practical values of the minimum local cycles length $g$ and aim at estimating the minimum syndrome former
constraint length which is needed to ensure that shorter cycles do not exist.
In the following we provide theoretical lower bounds of this type for $g = 6$ and $g = 8$.

\subsection{Absence of cycles with length $< g = 6$ \label{subsec:Cycles6} }

In order to meet the condition $g = 6$, we must ensure that local cycles with length $4$ do not exist.
Such short cycles occur when, for some $i,j,i',j'$, $j \neq j'$,
\begin{equation}
\delta_{i,j} = \delta_{i',j'} \quad \textrm{and} \quad l_s = l_s',
\label{eq:ConditionCycles4}
\end{equation}
\textit{i.e.}, in order to avoid length $4$ cycles there must not be any two equal differences starting from the same level.
We observe that the two differences may even be in the same row of $\HH_s$.

Let us first consider a regular $\HH_s$ with row weight $w=2$.
In this case, each row of $\HH_s$ only contains one difference $\delta_{i,j}$ and each difference can be used up to $c$ times without incurring length $4$ cycles (by using all the possible $c$ levels as starting levels).
For a given $L_h$, the differences starting from the first one of the $c$ available levels can take up to $L_h-1$ values.
Similarly, the differences starting from the second level can take up to $L_h-2$ values, and so on, until up to $L_h-c$ values
for the differences starting from the last level.
Since the differences corresponding to any two of the $a$ rows of $\HH_s$ must be different
in value and/or starting level, we have 
\begin{equation}
a \le  \sum_{i=0}^{c-1} (L_h - i -1) = c L_h - {c+1 \choose 2},
\label{eq:aupperboundCycles4}
\end{equation}
that is 
\[
L_h \ge \Bigg\lceil \frac{a+{c+1 \choose 2}}{c} \Bigg\rceil.
\]
Considering that it must be $L_h > c$, we have
\begin{equation}
L_h \ge \max \Bigg\{c+1, \Bigg\lceil \frac{a+{c+1 \choose 2}}{c} \Bigg\rceil \Bigg\}.
\label{eq:LhBoundw2Cycles4}
\end{equation}

We can extend \eqref{eq:aupperboundCycles4} to the case of a regular $\HH_s$ with row weight $w > 2$
by considering that, in such a case, each row of $\HH_s$ corresponds to $w \choose 2$ differences that
must meet condition \eqref{eq:ConditionCycles4}.
Hence \eqref{eq:aupperboundCycles4} becomes
\[
a{w \choose 2} \le c L_h - {c+1 \choose 2},
\]
while \eqref{eq:LhBoundw2Cycles4} becomes
\begin{equation}
L_h \ge \max \Bigg\{c+1, \Bigg\lceil \frac{a {w \choose 2}+{c+1 \choose 2}}{c} \Bigg\rceil \Bigg\}.
\label{eq:LhBoundCycles4Regular}
\end{equation}

When we have an irregular $\HH_s$ with row weights $w_i$, $i=0,1,2,\ldots, a-1$, each row of $\HH_s$ 
corresponds to $w_i \choose 2$ differences. Therefore \eqref{eq:LhBoundCycles4Regular} becomes
\begin{equation}
L_h \ge \max \Bigg\{c+1, \Bigg\lceil \frac{ \sum_{i=0}^{a-1}{w_i \choose 2}+{c+1 \choose 2}}{c} \Bigg\rceil \Bigg\}.
\label{eq:LhBoundCycles4Irregular}
\end{equation}

\subsection{Absence of cycles with length $< g = 8$ \label{subsec:Cycles8} }

The minimum length of local cycles is $g = 8$ when condition \eqref{eq:ConditionCycles4} is met and
length $6$ cycles of the type shown in Fig. \ref{fig:CycleExample} and described in Section \ref{sec:LocalCycles}
are avoided.

Let us first consider the case with $c=1$ and $\HH_s$ with row weight $w=2$.
Since summing two odd integers we always get an even number, the following proposition easily follows.
\begin{Pro}
For $c=1$ and $w=2$, if all the $\delta_{i,j}$ are different and odd, then local cycles with length $< g = 8$ do not exist.
\label{luno}
\end{Pro}
From Proposition \ref{luno} it follows that, if we wish to minimize $L_h$, we can choose the values of $\delta_{i,j}$ equal to $\left\{1,3,5,\ldots,2a-1\right\}$ and the code will be free of cycles with length $< g = 8$.

Another possible choice yielding absence of cycles with length $< g = 8$ follows from the fact that, for a given odd integer $x$, summing two values $\in \left[\frac{x+1}{2}; x\right]$ always gives a result $> x$.
Therefore, the following proposition holds.
\begin{Pro}
For $c=1$ and $w=2$, if the $\delta_{i,j}$ values are equal to $\left\{a, a+1, a+2, \ldots, 2a-1\right\}$, then local cycles with length $< g = 8$ do not exist.
\label{ldue}
\end{Pro}
Based on these propositions, we can prove the following lemma.
\begin{Lem}
For $c=1$ and $w=2$, local cycles with length $< g = 8$ can be avoided iff
\begin{equation}
L_h \ge 2a.
\label{eq:LhBoundc1w2Cycles8}
\end{equation}
\label{lem:LhBoundc1w2Cycles8}
\end{Lem}
\begin{IEEEproof}
From Propositions \ref{luno} and \ref{ldue} we have that the maximum value of a difference that is needed to avoid cycles with length $< g = 8$ is $2a-1$.
Therefore we have $L_h \ge 1+2a-1=2a$.
In order to prove the converse, let us consider that, for a given even integer $y$, summing two values $\in \left[1;\frac{y}{2}\right]$ always gives a result $\in \left[\frac{y}{2}+1; y\right]$.
In general, from the set $\left[1;y\right]$ we can select at most $\frac{y}{2}$ values which may be summed pairwise resulting in other values in the same set.
If we choose the values of the differences from the set $\left[1; 2a - 2\right]$, we only have $a - 1$ values which may be summed pairwise resulting in other values in the same set.
Therefore, we can only allocate $a - 1$ differences without introducing length $6$ cycles, which is not sufficient to cover all the $a$ rows of $\HH_s$.
\end{IEEEproof}

Equation \eqref{eq:LhBoundc1w2Cycles8} can be extended to the case $c > 1$ by considering that, in such a case, each difference value can be repeated up to $c$ times (by exploiting all the $c$
available levels as starting levels). Therefore, for $w = 2$ and $c > 1$ we have
\begin{equation}
L_h \ge \max \left\{c+1, \frac{2a}{c} \right\}.
\label{eq:LhBoundw2Cycles8}
\end{equation}

Let us consider larger values of $w$, \textit{i.e.}, $w \ge 3$.
For $c=1$, each row of $\HH_s$ has one or more cycles with length $6$, since at least $3$ symbols $1$ are at the same level (as described in Section \ref{sec:LocalCycles}).
Instead, for $w \ge 3$ and $c>1$ we can follow the same approach used for the case with $g = 6$, thus obtaining
\begin{equation}
L_h \ge \max \left\{c+1, \frac{2a {w \choose 2}}{c} \right\}.
\label{eq:LhBoundCycles8}
\end{equation}
 
When the rows of $\HH_s$ are irregular with weights $w_i$, $i=0,1,2,\ldots, a-1$, as done for the case with $g = 6$, we can consider that each row of $\HH_s$ corresponds to $w_i \choose 2$
differences and hence \eqref{eq:LhBoundCycles8} becomes
\begin{equation}
L_h \ge \max \Bigg\{c+1, \Bigg\lceil \frac{2 \sum_{i=0}^{a-1}{w_i \choose 2}}{c} \Bigg\rceil \Bigg\}.
\label{eq:LhBoundCycles8Irreg}
\end{equation}

\section{Examples \label{sec:Examples} }

\begin{figure}[!t]
\begin{centering}
\includegraphics[width=90mm,keepaspectratio]{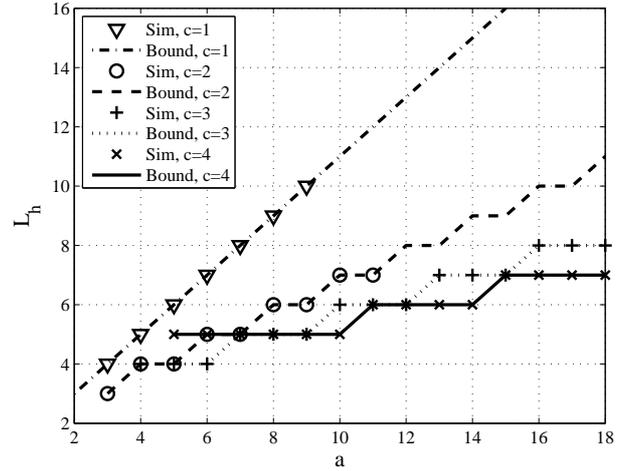}
\caption{Bounds on $L_h$ and values found through exhaustive searches as a function of $a$, for $w=2$, $g=6$ and some values of $c$.}
\label{fig:w2c1234g6}
\par\end{centering}
\end{figure}

\begin{figure}[!t]
\begin{centering}
\includegraphics[width=90mm,keepaspectratio]{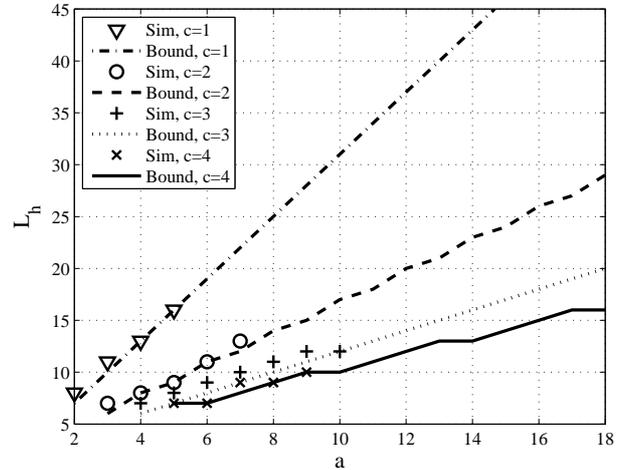}
\caption{Bounds on $L_h$ and values found through exhaustive searches as a function of $a$, for $w=3$, $g=6$ and some values of $c$.}
\label{fig:w3c1234g6}
\par\end{centering}
\end{figure}

\begin{figure}[!t]
\begin{centering}
\includegraphics[width=90mm,keepaspectratio]{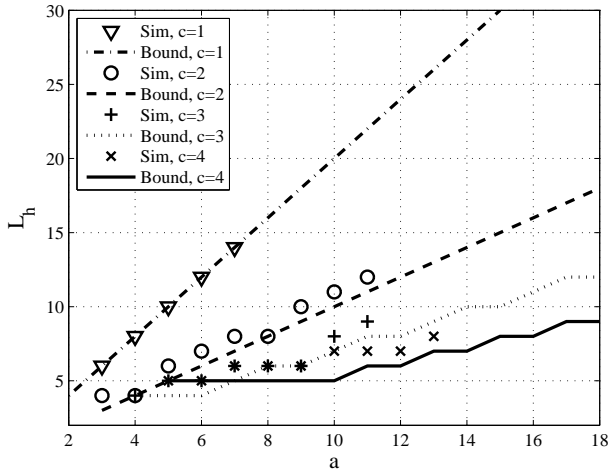}
\caption{Bounds on $L_h$ and values found through exhaustive searches as a function of $a$, for $w=2$, $g=8$ and some values of $c$.}
\label{fig:w2c1234g8}
\par\end{centering}
\end{figure}

In Figs. \ref{fig:w2c1234g6}-\ref{fig:w2c1234g8} we report the bounds on $L_h$ obtained as described in Section \ref{sec:MinConstLength}
as a function of $a$, for some values of $w$, $g$ and $c$.
We also compare these bounds with the results obtained through exhaustive searches over all the possible choices of $\HH_s$, performed through efficient numerical tools.

From Fig. \ref{fig:w2c1234g6} we observe that, for the cases with $w=2$ and $g=6$, the matching between the theoretical bound and the values found through exhaustive searches
is perfect for all the considered values of $c$.
Indeed, in this situation, all the practical cases are modeled by the bound, therefore it is always possible to find a solution achieving the bound.
Instead, when we have larger row weights of $\HH_s$, the theoretical bound may not be achievable in practical terms.
This results from Fig. \ref{fig:w3c1234g6} for $w=3$. 
However, we also observe that the deviations of the experimental values from the theoretical curves are rather small.
The results of exhaustive searches are well matched with the theoretical bounds also for the case with $w=2$ and $g=8$, as we observe from Fig. \ref{fig:w2c1234g8}.
In this case, we note that the gap to the bound increases for increasing values of $c$.

The same efficient tools used to perform exhaustive searches can also be exploited to perform Montecarlo experiments aimed at finding codes with small constraint length and absence of cycles with length up to some value $g$.
This way, it has been possible to find improved results with respect to previous solutions from the constraint length standpoint.
For example, in \cite{Bocharova2012} a code with $a=6$, $c=3$, $w=3$ and $g=10$ is provided with
\[ H(x)=\left[ \begin{array}{cccccc}
1 & x^{2} & x^{24} & x^{25} & x^{54} & x^{85}  \\
1 & x^{21} & x^{15} & x^{11} & x^{8} & x^{59} \\
1 & 1 & 1 & 1 & 1 & 1 \end{array} \right],
\] 
having $m_h = 85$ and $L_h = 258$.
Through a Montecarlo search performed with the tools described above, we have found a code with the same parameters and girth, having
\[
H(x)=\left[ \begin{array}{cccccc}
1 & x^{33} & 1 & x^{17} & x^{30} & x^{11}  \\
x^{16} & x^{8} & x^{33} & 1 & 1 & x^{33}  \\
x^{38} & 1 & x^{34} & x^{20} & x^{4} & 1 \end{array} \right],
\] 
\textit{i.e.}, $m_h = 38$ and $L_h = 117$, thus resulting in a considerable reduction over the former.
Similarly, in \cite{Zhou2012} a code with $a=5$, $c=3$, $w=3$ and $g=12$ is provided with
\[
H(x)=\left[ \begin{array}{cccccc}
x^{166} & x^{181} & x^{19} & 1 & x^{58} \\
x^{12} & x^{95} & 1 & x^{154} & x^{138}  \\
x^{27} & 1 & x^{185} & x^{117} & x^{170} \end{array} \right],
\] 
having $m_h = 185$ and $L_h = 558$, while we were able to find a code with the same parameters and girth, having
\[ 
H(x)=\left[ \begin{array}{cccccc}
x^{52} & 1 & x^{32} & 1 & x^{48} \\
1 & x^{51} & x^{47} & x^{45} & 1  \\
x^{33} & x^{25} & 1 & x^{16} & x^{44} \end{array} \right].
\] 
This code has $m_h = 52$ and $L_h = 159$, which also is a considerable improvement.
Another example in \cite{Zhou2012} with the same choice of the parameters achieves $m_h=134$, 
which still is considerably larger than the value we have found.

Concerning performance of these codes, there is a trade-off with their constraint length. However, codes with moderately small constraint lengths may still achieve better performance than their block counterparts. For example, we have verified through Montecarlo simulations of BPSK modulated transmission over the AWGN channel that one of our LDPCC codes with $w=3$, $a=9$, $c=3$, $g=8$ and $v_s=1143$ exhibits a gain of about $0.3$ dB at $\mathrm{BER}=10^{-5}$ with respect to the WiMax standard LDPC block code with the same rate ($2/3$) and length $2304$.

\section{Conclusion \label{sec:Conclusion}}

We have studied the design of time-invariant \ac{SC-LDPC} codes with small constraint length and
free of local cycles up to a given length.
By directly designing the syndrome former matrix, we have obtained codes with smaller constraint length
with respect to those designed by unwrapping \ac{QC-LDPC} block codes.
We have also provided theoretical lower bounds on the minimum constraint length which is needed to
achieve codes with a fixed minimum length of the local cycles, and shown through exhaustive searches
that practical codes achieving or, at least, approaching these bounds can be found.

\bibliographystyle{IEEEtran}
\bibliography{Archive}

\end{document}